\documentclass[nofootinbib,11pt,preprintnumbers,floatfix]{revtex4-2}
\usepackage[utf8]{inputenc}

\usepackage[dvipsnames]{xcolor}
\definecolor{red}{rgb}{0.9, 0,0}
\definecolor{cerulean}{rgb}{0., 0.42,0.9}
\definecolor{navy}{rgb}{0.05, 0.05,0.8}

\usepackage[colorlinks]{hyperref}
\hypersetup{
    colorlinks = true,
    citecolor  = red,
	linkcolor  = navy
}

\usepackage{slashed}
\usepackage{graphics}
\usepackage{amsmath}
\usepackage{amssymb}
\usepackage{latexsym}
\usepackage{mathtools}
\usepackage{dsfont}
\usepackage{hyperref}
\usepackage{amsfonts}
\usepackage{enumitem}
\usepackage{xstring} 
\usepackage{xspace} 
\usepackage[normalem]{ulem}
\usepackage{fontawesome}
\usepackage{subcaption}
\usepackage{braket}

\newcommand{\tr}{{\rm tr}}

\newcommand{\GHY}{{\rm GHY}}

\newcommand{\EH}{{\rm EH}}

\newcommand{\AdS}{{\rm AdS}}
\newcommand{\Mink}{{\rm Mink}}
\newcommand{\matter}{{\rm matter}}
\newcommand{\Poincare}{Poincar\'e\xspace}
\newcommand{\RT}{{\rm r.t.}}

\newcommand{\gt}{\tilde{g}}
\newcommand{\dt}{\tilde{\nabla}}
\newcommand{\bt}{\tilde{\Box}}
\newcommand{\rt}{\tilde{R}}
\newcommand{\mr}{\mathfrak{r}}

\usepackage{subfiles} % Best loaded last in the preamble

\begin{document}

\title{Near-Horizon Quantum Dynamics of 4-d Einstein Gravity from 2-d JT Gravity}

\author{Sergei Gukov, Vincent S. H. Lee, Kathryn M. Zurek}
\affiliation{Walter Burke Institute for Theoretical Physics, California Institute of Technology, Pasadena, CA}

\preprint{CALT-TH-2022-016}

\begin{abstract}
    We study quantum fluctuations in the lightcone metric of the 4-d Einstein-Hilbert action via dimensional reduction to Jackiw-Teitelboim (JT) gravity. In particular, we show that, in Einstein gravity, the causal development of a region in flat Minkowski spacetime, near a horizon defined by light sheets, can be described by an effective two-dimensional dilaton theory. This enables us to make use of known solutions of the JT action, where the spacetime position of a horizon  has quantum uncertainty due to metric fluctuations. This quantum uncertainty can be then directly related to the original 4-d light cone coordinates, allowing us to compute the uncertainty in the time of a photon to travel from tip-to-tip of a causal diamond in flat 4-d Minkowski space.  We find that both Planck and infrared scales (with the latter set by the size of the causal diamond) enter the uncertainty in photon travel time, such that the quantum fluctuation in the arrival time may be observably large.  
\end{abstract}

\maketitle
\newpage
\tableofcontents
\newpage

 %-------------------------------------------------------------------------------------------------
%
% 	         Introduction
% %-------------------------------------------------------------------------------------------------

\section{Introduction}
\label{sec:introduction}

Dimensional reduction has long played an important role in understanding the behavior of higher dimensional gravitational theories, in particular in the study of black hole horizons. When a higher-dimensional theory is reduced to a two-dimensional system associated with the light-cone directions, the area of the transverse directions becomes the dilaton field. In the near-horizon limit the dilaton is conformal (see {\em e.g.}~\cite{Solodukhin_1999}), and by studying the conformal states of the action and using Cardy's formula the correct expression for the black hole entropy can be derived.

While the conformal description of near-horizon states has been widely applied to black hole horizons, there is reason to think that a similar formalism may apply to light sheet horizons more generally~\cite{banks}.  The interior of a causal diamond in many generic spacetimes can be represented by a topological black hole metric:
\begin{equation}\label{eqn:causal_diamond_general}
	ds^2 = -f(R) dT^2 + \frac{dR^2}{f(R)} + \rho(T,R)^2 d\Sigma_{d-2}^2.
\end{equation}
For example, for boundary anchored diamonds in AdS, $f(R) = R^2/L^2 - 1$, with $L$ the AdS curvature, while in empty Minkowski, $f(R) = 1-R/R_h$, where $R_h$ is the radius of the bifurcate horizon. The representation of the causal diamond has an associated modular Hamiltonian, $K$, that characterizes the density matrix of the diamond, $\rho_{\mathrm{diamond}} = e^{-K}/\tr(e^{-K})$.  If one conjectures that the near-horizon states of a light-sheet horizon are described by a conformal field theory, one is able to immediately write down the form of the partition function (see discussion in~\cite{banks}), 
\begin{equation}
\log Z = \log \left(\int dE ~e^{B \sqrt{E} - \beta E}\right),
\end{equation}
from which one can derive both the expectation value of $K$, $\langle K \rangle = -\beta\partial_{\beta}\log Z+\log Z=\beta\langle E\rangle+\log Z=S$, and its fluctuations, $\langle \Delta K^2 \rangle =  \beta^2 \partial^2_\beta \log Z = \beta^2\langle \Delta E^2 \rangle$, finding $\langle \Delta K^2 \rangle = \langle K \rangle$.  This result agrees with previous calculations for Ryu-Takayanagi diamonds in AdS/CFT~\cite{de_Boer_2019,vz2}.  These modular fluctuations generate metric fluctuations, inducing a quantum uncertainty in the horizon of the causal diamond.  The authors of previous works~\cite{banks,vz2,vz1} suggested that these fluctuations might be observably large.

\begin{figure}
	\includegraphics[width=1.\textwidth]{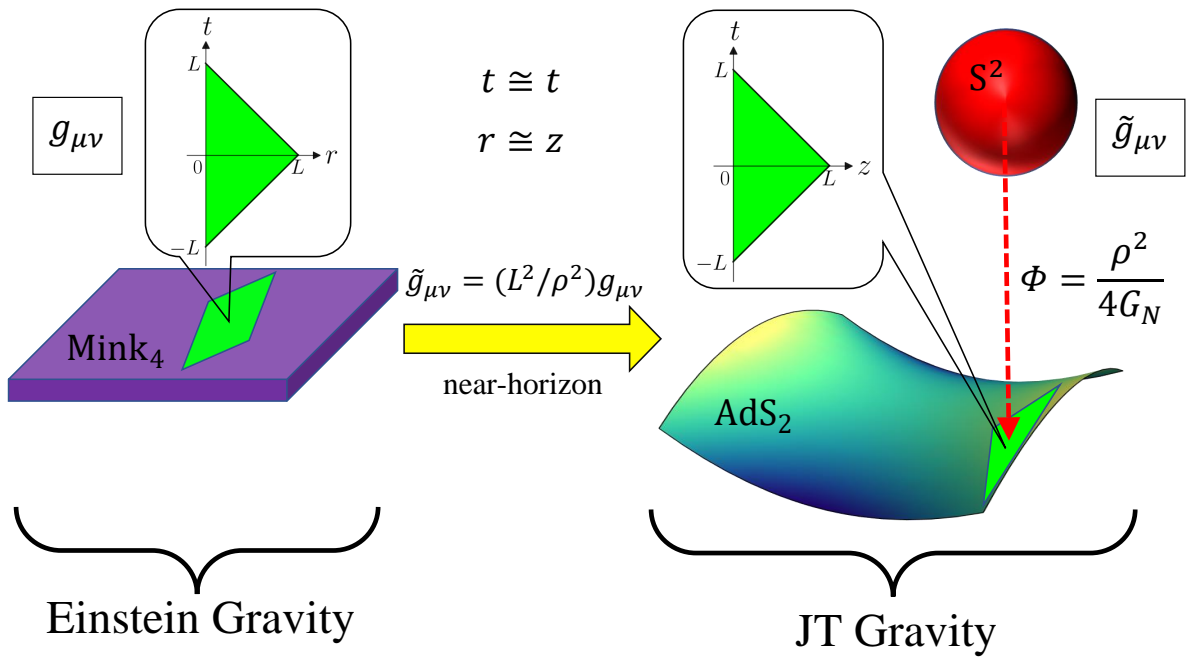} 
	\caption{The metric on Mink$_4$ is conformally equivalent to that on AdS$_2\times$S$^2$. Inserts illustrate spacetime diagrams of the causal diamonds in both geometries. }\label{fig:ads2s2}
\end{figure}

Here, we put a new twist on these ideas by showing that Einstein gravity on a causal diamond in flat 4-d spacetime, at least in the near-horizon limit, exactly dimensionally reduces to Jackiw-Teitelboim (JT) gravity~\cite{jackiw1985343, teitelboim198341} in 2 dimensions.  In particular, the parent 4-d theory can be Weyl-rescaled to a dilaton theory on AdS$_2\times$S$^2$, as shown in Fig.~\Ref{fig:ads2s2}. The dynamics of the dilaton (shown by a dashed line) controls both the size of the S$^2$ and the relative position of the horizon with respect to the boundary. As has also been noted by others, the dilaton is expected to have an effective hydrodynamic description~\cite{harlow2019factorization}.  

This implies that, if we are interested only in observables defined on a light sheet horizon, we can make use of a vast literature studying the JT theory.  In turn, this allows us to potentially draw connection between experimental observables and theoretical calculations in the vast field of quantum gravity.  We will make use, in particular, of the solutions presented in Ref.~\cite{harlow2019factorization}, which features a 2-sided AdS$_2$ spacetime.  These authors computed the quantum uncertainty in a geodesic distance controlled by the dilaton.  We will show that the quantum uncertainty in this geodesic distance computed in the 2-sided 2-d JT theory is directly related to the uncertainty in the travel time for a photon to be fired from a boundary to the bulk, reflected by a mirror, and returned to the boundary. The relation is illustrated in Fig.~\ref{fig:ads2s2}, which will be described in more detail in the main text.  
The original 4-d spacetime has a flat metric
\begin{equation}\label{eqn:Minkowski}
	ds^2_{\Mink} = -dt^2 + dr^2 + r^2d\Omega_2^2 \, ,
\end{equation} 
which is conformally equivalent to the product metric on AdS$_2\times$S$^2$: 
\begin{equation}\label{eqn:AdS2_S2}
	ds^2_{\Mink} = \frac{\rho^2}{L^2}\left(L^2\frac{-dt^2+dr^2}{r^2} + L^2d\Omega_2^2\right) \, ,
\end{equation} 
where we treat $\rho=r$ as a scalar field, and $L$ is a positive constant which we identify with the AdS$_2$ radius (as well as the radius of the sphere S$^2$). This closely resembles the near-horizon limit of a 4-d near-extremal Reissner-Nordstr\"{o}m black hole. In the limit $l_p\to 0$, such that the magnetic charge and temperature are kept fixed, this near-horizon geometry becomes AdS$_2\times$S$^2$, while the Einstein-Maxwell action reduces to the JT action~\cite{maldacena_Michelson_Strominger} after integrating over the angular coordinates. Moreover, Eq.~\eqref{eqn:AdS2_S2} demonstrates that the Minkowski spacetime variables, $t$ and $r$, naturally become the AdS spacetime variables in \Poincare coordinates, $t$ and $z$. Motivated by this well known result, in later sections we demonstrate that the Einstein gravity reduces to the JT gravity by a similar procedure and compute the relevant observables.

The causal diamond can be observationally defined by an interferometer set-up, also shown in Fig.~\ref{fig:ads2s2}. We align our interferometer arm along the radial direction and denote its length by $L$. The photon is fired from $r=0$ at $t=-L$. At time $t=0$, the photon hits the mirror at the interferometer end ($r=L$) and bounces back. Finally, the photon arrives to its starting position $r=0$ at $t=L$. The photon trajectory is hence described by $t-r=-L$ and $t+r=L$ for its first and second trip respectively. The spacetime region bounded by the photon trajectory and $r=0$ is commonly referred to as the causal diamond, where the photon trajectory itself is known as the horizon since no particles can cross the boundary and escape to infinity, similar to a black hole horizon. Spacetime fluctuations lead to uncertainty in the photon travel time.

At first sight, it seems surprising that JT gravity in 2-sided AdS, such as that solved explicitly in Ref.~\cite{harlow2019factorization}, would have anything to do with photon trajectories in Minkowski space.  After all, the JT set-up is in AdS space and has curvature, while Minkowski does not.  It is also not immediately clear what a 2-sided geometry has to do with a 1-sided causal diamond. It is important to highlight two subtle, but important points that allow us to utilize the computational tools of JT gravity.   First, while AdS clearly has curvature that the parent Minkowski theory does not, in the near horizon limit of the causal diamond relevant for photon trajectories, we will show that these terms in the action which contribute to the curvature are subdominant.  Second, while the formal solution to JT gravity that we utilize is in two-sided AdS,
%First, as we will demonstrate in the next section, the dimensional reduction from Einstein gravity to JT gravity is only valid near the casual diamond horizon. As such, if one directly inserts the equation of motion of the scalar field $\rho$ derived from the JT action, obtained formally by truncating the dimensionally reduced Einstein-Hilbert action, back to the parent metric in Eq.~\eqref{eqn:AdS2_S2}, one will obtain an unphysical curvature. This however will not affect the observable of an interferometer experiment as the photon travels along the causal diamond horizon, the region where truncating the action is justified, as we will show in the next section. Second, the formal solution to JT gravity is a two-sided AdS. However, the physical system, which is a causal diamond in the parent Minkowski spacetime, only covers one side of the AdS after dimensional reduction. Utilizing the global coordinate system in AdS, 
we will see that the length of the causal diamond horizon in one-sided Minkowski is identical to that of the two-sided AdS horizon connecting the two boundaries. Hence, one can compute the physical photon roundtrip time by calculating the time for a photon to travel from one AdS boundary to the boundary on the other side of AdS along the horizon. We will see that the other side of AdS serves as a convenient tool for us to compute quantum fluctuations in the physical observable by computing how one side of AdS fluctuates with respect to the other side.

The outline of the paper is as follows. In Sec.~\ref{sec:jt_derivation}, we discuss how the JT action can be obtained by dimensionally reducing the familiar gravitational action and dropping a subdominant kinetic term in the near-horizon limit. In Sec.~\ref{sec:AdS}, we study the AdS geometry and introduce various useful coordinate systems. In Sec.~\ref{sec:spacetime_fluc_JT}, we define our observable in the context of JT gravity and compute its fluctuations. Finally, in Sec.~\ref{sec:conclusion}, we discuss implications of our results and mention a few future directions.

%-------------------------------------------------------------------------------------------------
%
% 	         Derivation of the JT action 
% %-------------------------------------------------------------------------------------------------

\section{Dimensional Reduction to the JT action}
\label{sec:jt_derivation}

We begin by dimensionally reducing the familiar gravitational action in 4-d Minkowski spacetime, in the near-horizon limit, to the 2-d JT action. As advertised above, this calculation is similar to the previous work by one of the authors~\cite{banks} on small empty diamonds.
There are, however, a couple of important differences with these earlier works. First, in line with theories of JT gravity, our dimensionally-reduced manifold has a boundary.  We thus must include the boundary contributions during the dimensional reduction process, which will ultimately lead to the boundary action in JT gravity. This is crucial for our later analysis since the bulk action vanishes on-shell for JT gravity, and thus the boundary term gives rise to the sole degree of freedom. Second, we perform a (different) Weyl rescaling to bring the two-dimensional metric into the AdS$_2$ form to align with the exact JT gravity setup studied in the literature.  

On a 4-manifold $M_4$, the total action, $I=I_{\EH}+I_{\GHY}$, is the sum of the bulk Einstein-Hilbert (EH) action and the boundary Gibbons-Hawking-York (GHY) action:
\begin{align}\label{eqn:I_EH_GHY}
	I_{\EH} &= \frac{1}{16\pi G_N} \int_{M_4} d^4x \sqrt{-g_4} R_4 \nonumber \\
	I_{\GHY} &= \frac{1}{8\pi G_N} \int_{\partial M_4} d^3x \sqrt{-\gamma_3} K_3 \, 
\end{align} 
where $G_N$ is the 4-d gravitational constant, $\gamma_{3}$ is the induced metric on the boundary, $g_4$ is the metric with the Ricci scalar $R_4$ and the extrinsic curvature $K_3$ on $\partial M_4$. The GHY action is needed in gravitational theories with a boundary to make the variational problem well-posed. In particular, the extra boundary term that arises from varying the EH action cancels against the variation of the GHY term. We will see a similar mechanism in action shortly.

We consider spherically-symmetric metrics in the general form
\begin{equation}\label{eqn:G_N}
	ds^2 = g_{ab}(x^0,x^1)dx^adx^b + \rho^2(x^0,x^1)d\Omega_2^2 \, .
\end{equation}
where $x^0$ and $x^1$ will be referred to as the light-cone coordinates\footnote{We use Greek letters for bulk coordinates in four dimensions and Latin letters from the early part of the alphabet for the light-cone coordinates.}, the radius $\rho$ is a scalar function of $x^0$ and $x^1$, and $d\Omega_2^2$ is the line element of a two-dimensional unit sphere. Geometrically speaking, $\rho$ sets the radius of the horizon. As we will see below, $\rho^2$ plays the role of a dilaton, which corresponds to the horizon area (and hence the entropy).

A generalization of the conformal equivalence Mink$_4\cong$AdS$_2\times$S$^2$ noted in Eqs.~\eqref{eqn:Minkowski}-\eqref{eqn:AdS2_S2} is a similar relation between a spherically-symmetric metric~\eqref{eqn:G_N} and the space of the form $\tilde{M}_2\times$S$^2$:
\begin{equation}\label{eqn:M2_S2}
	ds^2 = \frac{\rho^2}{L^2}\left(\frac{L^2}{\rho^2}g_{ab}dx^adx^b + L^2d\Omega_2^2\right) \, .
\end{equation} 
Ultimately we would like to work with an $\AdS_2$ metric, which motivates us to denote the metric in the parenthesis as $\tilde{g}_{\mu\nu}=(L^2/\rho^2)g_{\mu\nu}$, and compute the action in terms of $\gt_{\mu\nu}$. 

A few remarks are in order:
\begin{itemize}
	\item Since Einstein gravity is not conformally invariant, $\gt_{\mu\nu}$ does not satisfy the usual vacuum Einstein equation. However, it still satisfies the equation of motion that follows from action Eq.~\eqref{eqn:I_EH_GHY} after the contribution of the conformal factor is properly accounted for. 
	\item The conformal relation between $g_{\mu\nu}$ and $\gt_{\mu\nu}$ in Eq.~\eqref{eqn:AdS2_S2} works for any choice of positive $L$. We find it most convenient to choose $L$ that coincides with the interferometer arm length.
	\item Weyl transformations do not alter the causal structure of a metric. A null geodesic in $g_{\mu\nu}$ is still a null geodesic in $\gt_{\mu\nu}$. 
\end{itemize}

\subsection{Einstein-Hilbert Action}
We first consider the EH action. The curvatures of $g_{\mu\nu}$ and $\gt_{\mu\nu}$ are related by~\cite{carroll_2019}
\begin{equation}\label{eqn:R4}
	R_4 = L^2(\rho^{-2}\tilde{R}_4-6\rho^{-3}\bt\rho) \, ,
\end{equation} 
while the curvature of the product manifold $\gt_{\mu\nu}$ in Eq.~\eqref{eqn:M2_S2} is a simple sum of individual curvatures
\begin{equation}\label{eqn:R4_tilde}
	\tilde{R}_4 = \tilde{R}_2 +\frac{2}{L^2} \, ,
\end{equation} 
where $\tilde{R}_2$ is the Ricci scalar of $\tilde{g}_{ab}$. This allows us to evaluate the action in Eq.~\eqref{eqn:I_EH_GHY}\footnote{We use the shorthands $\Box\rho=g^{\mu\nu}\nabla_{\mu}\nabla_{\nu}\rho$, $(\nabla\rho)^2=g^{\mu\nu}\nabla_{\mu}\rho\nabla_{\nu}\rho$, $\bt\rho=\gt^{\mu\nu}\dt_{\mu}\dt_{\nu}\rho$ and $(\dt\rho)^2=\gt^{\mu\nu}\dt_{\mu}\rho\dt_{\nu}\rho$. Since $\rho$ does not depend on the angular variables, we can also replace the four-dimensional contractions in the above derivatives by just two-dimensional contractions.}
\begin{align}\label{eqn:4D_EH_conformal}
	I_{\EH} &= \frac{1}{16\pi G_N}\frac{1}{L^2 }\int_{\tilde{M}_4} d^4x \sqrt{-\gt_4} \left(\rho^2\rt_2 - 6\rho\bt\rho + \frac{2}{L^2}\rho^2\right) \nonumber \\
	&= \frac{1}{16\pi G_N}\frac{1}{L^2 }\int_{\tilde{M}_4} d^4x \sqrt{-\gt_4} \left(\rho^2\rt_2 + 6(\dt\rho)^2+ \frac{2}{L^2}\rho^2\right) - \frac{1}{16\pi G_N}\frac{3}{L^2} \int_{\partial\tilde{M}_4} d^3x \sqrt{-\tilde{\gamma}_3}\gt^{\mu\nu}\tilde{n}_{\mu}\dt_{\nu}\rho^2 \, .
\end{align} 
The boundary term here comes from the Stokes' theorem\footnote{We define the normal vector $n^{\mu}$ to the boundary to be pointing outward/inward if it is spacelike/timelike. In this convention, the Stokes' theorem reads $\int_{M_4} d^4x \sqrt{-g_4} \nabla_{\mu}V^{\mu} = \int_{\partial M_4} d^3x \sqrt{-\gamma_3} n_{\mu}V^{\mu}$ for any vector $V^{\mu}$ regardless of the signature of the boundary. The analogous formula also holds in two dimensions.} that relates a total derivative to a boundary term.

To perform the dimensional reduction, we integrate over the angular directions while keeping in mind that $\rho$ as well as $\tilde{R}_2$ only depend on the light-cone variables. Hence, the EH action becomes
\begin{align}\label{eqn:I_EH_d_3}
	I_{\EH} &= \frac{1}{4G_N}\int_{\tilde{M}_2} d^2x \sqrt{-\gt_2} \left(\rho^2\rt_2 + 6(\dt\rho)^2+ \frac{2}{L^2}\rho^2\right) - \frac{3}{4G_N} \int_{\partial\tilde{M}_2} dx^0 \sqrt{-\tilde{\gamma}_1}\gt^{ab}\tilde{n}_a\dt_b\rho^2 \, ,
\end{align} 
where $x^0$ is the boundary time.

\subsection{Gibbons-Hawking-York Action}
We now turn our attention to the GHY action. The normal vector of the boundary transforms as $\tilde{n}^{\mu}=(\rho/L)n^{\mu}$, hence the extrinsic curvature transforms as
\begin{align}\label{eqn:K_tilde}
	K_3 &= \nabla_{\mu}n^{\mu} \nonumber \\
	&= \frac{1}{\sqrt{-g_4}}\partial_{\mu}(\sqrt{-g_4}n^{\mu}) \nonumber \\
	&= \left(\frac{L}{\rho}\right)^4\frac{1}{\sqrt{-\gt_4}}\partial_{\mu}\left(\left(\frac{\rho}{L}\right)^3\sqrt{-\gt_4}\tilde{n}^{\mu}\right) \nonumber \\
	&= \frac{L}{\rho}\tilde{K}_3 +3\frac{L^2}{\rho^2}\tilde{n}^{\mu}\dt_{\mu}\frac{\rho}{L} \, .
\end{align} 
Putting this into the GHY action in Eq.~\eqref{eqn:I_EH_GHY} gives
\begin{equation}\label{eqn:4D_GHY_conformal}
	I_{\GHY} = \frac{1}{8\pi G_N}\frac{1}{L^2}\int_{\partial \tilde{M}_4} d^3x \sqrt{-\tilde{\gamma}_3} \left(\rho^2\tilde{K}_3 + \frac{3}{2}\gt^{\mu\nu}\tilde{n}_{\mu}\dt_{\nu}\rho^2\right) \, .
\end{equation} 
Since the boundary $\partial M_4$ is taken to be spherically symmetric, only the light-cone component of the normal vector $n^{\mu}$ is non-zero, which then coincides with $n^a$, the normal vector to $\partial M_2$. On the other hand, projection to $\tilde{M}_2$ gives a simple relation $\tilde{K}_3=\tilde{K}_1$, where $\tilde{K}_1$ is the extrinsic curvature of $\gt_{ab}$ on $\partial\tilde{M}_2$. This allows us to perform the dimensional reduction
\begin{equation}\label{eqn:4D_GHY_conformal_final}
	I_{\GHY} = \frac{1}{2G_N}\int_{\partial\tilde{M}_2} dx^0 \sqrt{-\tilde{\gamma}_1} \rho^2\tilde{K}_1 + \frac{3}{4G_N}\int_{\partial\tilde{M}_2} dx^0 \sqrt{-\tilde{\gamma}_1}\gt^{ab}\tilde{n}_a\dt_b\rho^2 \, .
\end{equation} 
We see that the extra boundary term from the EH action precisely cancels the second term in the GHY action. The total action then becomes
\begin{equation}\label{eqn:I_total}
	I = \frac{1}{4G_N}\int_{\tilde{M}_2} d^2x \sqrt{-\gt_2} \left(\rho^2\rt_2 + 6(\dt\rho)^2+ \frac{2}{L^2}\rho^2\right) + \frac{1}{2G_N}\int_{\partial\tilde{M}_2} dx^0 \sqrt{-\tilde{\gamma}_1} \rho^2\tilde{K}_1 \, .
\end{equation} 
Similar cancellations have been noted in Ref.~\cite{Svesko:2022txo} while models with actions similar to Eq.~\eqref{eqn:I_total} have been extensively studied in Ref.~\cite{almheiri2015models}.

\subsection{Near-horizon Limit}

We now examine the metric and the action near the horizon of a Minkowski causal diamond of size $L$. The metric in the interior of a causal diamond is obtained from Eq.~\eqref{eqn:Minkowski} via the transformation~\cite{vz1}
\begin{align}\label{eqn:VZ1_transformation}
	t &= 2L\sinh\left(\frac{T}{2L}\right)\sqrt{1-\frac{R}{L}} \nonumber \\
	r &= L - 2L\cosh\left(\frac{T}{2L}\right)\sqrt{1-\frac{R}{L}} \, ,
\end{align} 
and the metric can be written in the form of Eq.~\eqref{eqn:causal_diamond_general}
\begin{equation}\label{eqn:VZ1_metric}
	ds_{\Mink}^2 = -\left(1-\frac{R}{L}\right)dT^2 + \frac{dR^2}{1-R/L} + \rho^2(T,R)d\Omega_2^2 \, ,
\end{equation} 
where we again identify $\rho=r$. The transformed light-cone variables are $T$ and $R$. Observe that
\begin{align}\label{eqn:VZ1_transformation_horizon}
	(t+r-L)(t-r+L) = -4L^2\left(1-\frac{R}{L}\right) \, ,
\end{align} 
hence the horizon of the causal diamond described at the end of Sec.~\ref{sec:introduction} is located at $R=L$. In the near-horizon limit, $R\to L$, the dilaton is approximately a large positive constant. We can thus expand the dilaton as a small perturbation
\begin{equation}\label{eqn:dilaton_perturbation}
	\rho^2 = \phi_0 + \phi \, ,
\end{equation} 
where $\phi_0=L^2$ and $\phi\ll \phi_0$. It is also clear in this coordinate that the classical area of the causal diamond is $A=4\pi L^2$.

The action we obtained in Eq.~\eqref{eqn:I_total} is almost the action of JT gravity except for the kinetic term $(\dt\rho)^2$. It has been argued in Ref.~\cite{Sarosi:2017ykf} that the kinetic term is a subdominant contribution in the context of a near-extremal Reissner-Nordstr\"{o}m black hole. We briefly review the argument and apply it to our set-up. Expanding the kinetic term using Eq.~\eqref{eqn:dilaton_perturbation} gives
\begin{equation}\label{eqn:kinetic_expanded}
	\int_{\tilde{M}_2}d^2x\sqrt{-\gt_2}(\dt\rho)^2 = \frac{1}{4}\int_{\tilde{M}_2}d^2x\sqrt{-\gt_2}\frac{(\dt\phi)^2}{\phi_0+\phi} \, .
\end{equation} 
Suppose the system is perturbed by coupling to some matter field via $I\to I + I_{\matter}$. Then, the equation of motion associated with the action in Eq.~\eqref{eqn:I_total} can be written as $T_{ab}=T^{\matter}_{ab}$ with $T_{ab}=-\dt_a\dt_b\rho^2$, where we have absorbed the dilaton kinetic term into the definition of $T^{\matter}_{ab}$. In the conformal gauge, $d\tilde{s}^2=-\exp(2\omega(u^+,u^-))du^+du^-$, the ++ component of the equation of motion turns out to be 
\begin{equation}\label{eqn:conformal_gauge_++}
	-e^{2\omega}\partial_+(e^{-2\omega}\partial_+\rho^2) = T^{\matter}_{++} > 0 \, .
\end{equation} 
Integrating this expression along a line $u^-=0$ from $u^+=0$ to $u^+=\pi$ then gives
\begin{equation}\label{eqn:integrate_Tmatter}
	\int_0^{\pi} du^+ e^{-2\omega}T^{\matter}_{++} = [e^{-2\omega}\partial_+\rho^2]|_{u^+\to 0} - [e^{-2\omega}\partial_+\rho^2]|_{u^+\to \pi} \, .
\end{equation} 
In $\AdS_2$, the conformal gauge is given by $\exp(-2\omega)\sim\sin^2u^+$ with the boundaries located at $u^+=0$ and $u^+=\pi$. Requiring the expression in Eq.~\eqref{eqn:integrate_Tmatter} be positive then implies that $\rho^2$ diverges near at least one of the boundaries~\cite{maldacena_Michelson_Strominger}
\begin{align}\label{eqn:dilaton_diverges}
	\rho^2|_{u^+\to 0} &\sim \mathrm{constant} + \frac{1}{u^+} \nonumber \\
	\rho^2|_{u^+\to \pi} &\sim \mathrm{constant} + \frac{1}{u^+-\pi} \, .
\end{align} 
With this information in hand, we can consider the dilaton kinetic term in Eq.~\eqref{eqn:kinetic_expanded} using the Poincar\'e coordinates\footnote{We use the symbol $t$ for both the Minkowski time and the AdS time in Poincar\'e coordinates since they can be identified with each other via the Weyl rescaling of Eq.~\eqref{eqn:AdS2_S2}, while different notations are used for the spatial coordinates, $z$ and $r$.}
\begin{equation}\label{eqn:poincare}
	d\tilde{s}^2 = L^2\frac{-dt^2+dz^2}{z^2} \, ,
\end{equation} 
where the boundary is located at $z=0$. Since the dilaton diverges as $\phi\sim 1/z$ near $z=0$ and has the dimension of $[\mathrm{length}]^2$, by dimensional analysis, one finds $\phi\sim l_p^2L^2E/z$ where $E$ is the energy associated with the causal diamond. The derivatives evaluate to $(\dt\phi)^2=g^{zz}\partial_z\phi\partial_z\phi=\phi^2/L^2\sim \phi^2/\phi_0$. Hence we can evaluate Eq.~\eqref{eqn:kinetic_expanded} 
\begin{align}\label{eqn:kinetic_term_small}
	\frac{(\dt\phi)^2}{\phi_0+\phi} &\approx \frac{1}{\phi_0}\frac{1}{1+\phi/\phi_0}\frac{\phi^2}{\phi_0} \nonumber \\
	&= \frac{\phi^2}{\phi_0^2} + \mathcal{O}\left(\frac{\phi^3}{\phi_0^3}\right) \, ,
\end{align} 
which is quadratic in $\phi/\phi_0$ at the leading order, and thus can be omitted in Eq.~\eqref{eqn:I_total}. This leaves us with the JT action
\begin{equation}\label{eqn:JT_def}
	I = \int_{\tilde{M}_2} d^2x\sqrt{-\gt_2}\Phi\left(\rt_2+\frac{2}{L^2}\right)+2\int_{\partial\tilde{M}_2} dx^0 \sqrt{-\tilde{\gamma}_1} \Phi\tilde{K}_1 \, .
\end{equation}
where we have defined the dimensionless dilaton field 
\begin{equation}\label{eqn:field_def}
	\Phi=\frac{\rho^2}{4G_N},
\end{equation} 
which controls the size of the S$^2$.  We will also show that this field controls how long it takes for a photon to traverse from the bottom to the top of the causal diamond.

We emphasize that the procedure of dropping the dilaton kinetic term (and hence the correspondence with JT gravity) is only valid near the causal diamond horizon. The classical equations of motion for the metric and dilaton are later derived in Eq.~\eqref{eqn:Poincare_coodinates} using the truncated action Eq.~\eqref{eqn:JT_def}. If one attempts to directly compute the classical Ricci scalar of the four dimensional metric in Eq.~\eqref{eqn:G_N} and Eq.~\eqref{eqn:field_def}, one finds a non-vanishing curvature for $r>0$. On the other hand, if one retains the kinetic term, than the metric equation of motion would remain the same as Eq.~\eqref{eqn:Poincare_coodinates}, but the dilaton solution would simply be $\rho=z=r$, which would completely reproduce the original four dimensional Minkowski metric in Eq.~\eqref{eqn:G_N} with zero curvature. For our purposes, we are interested in the dilaton equation of motion near a null trajectory, where the dilaton kinetic term is subdominant.  Dropping the kinetic term, however, comes at a price of introducing a (unphysical) curvature in the parent Minkowski theory.  According to our argument above, however, this curvature is irrelevant for the dilaton equations of motion in the near-horizon limit.  We thus proceed with the JT theory, in the near-horizon limit, as a good approximation to near-horizon Minkowski spacetime fluctuations.

%-------------------------------------------------------------------------------------------------
%
% 	         The two-sided AdS geometry and classical dilaton solution
% %-------------------------------------------------------------------------------------------------

\section{The Two-sided AdS Geometry and Classical Dilaton Solution}
\label{sec:AdS}
Before considering the quantum fluctuations, we discuss the classical equations of motion for both the metric field and the dilaton.  This will allow us to determine how the dilaton is related to fluctuations in geodesic distances, that we can in turn relate to photon travel times in the original 4-d Minkowski space.  The equations of motion read:
\begin{align}\label{eqn:classical_EOM}
	\rt_2 + \frac{2}{L^2} &= 0 \nonumber \\
	(L^2\dt_a\dt_b-\gt_{ab})\Phi &= 0 \, .
\end{align}
The first equation shows that the bulk geometry is a slice of AdS while the second equation specifies the classical behavior of the dilaton. To ensure that the variational problem is well defined, we fix the dilaton value at the boundary to be 
\begin{equation}
	\Phi|_{\mathrm{boundary}} = \frac{\Phi_b\mr_c}{L} \, ,
\end{equation} 
and the induced metric to be $\gamma_{00}|_{\partial\tilde{M}_2}=\mr_c^2/L^2$, where $\mr_c\to\infty$ is the regularized location of the AdS boundary. 

\begin{figure}
	\includegraphics[scale=0.8]{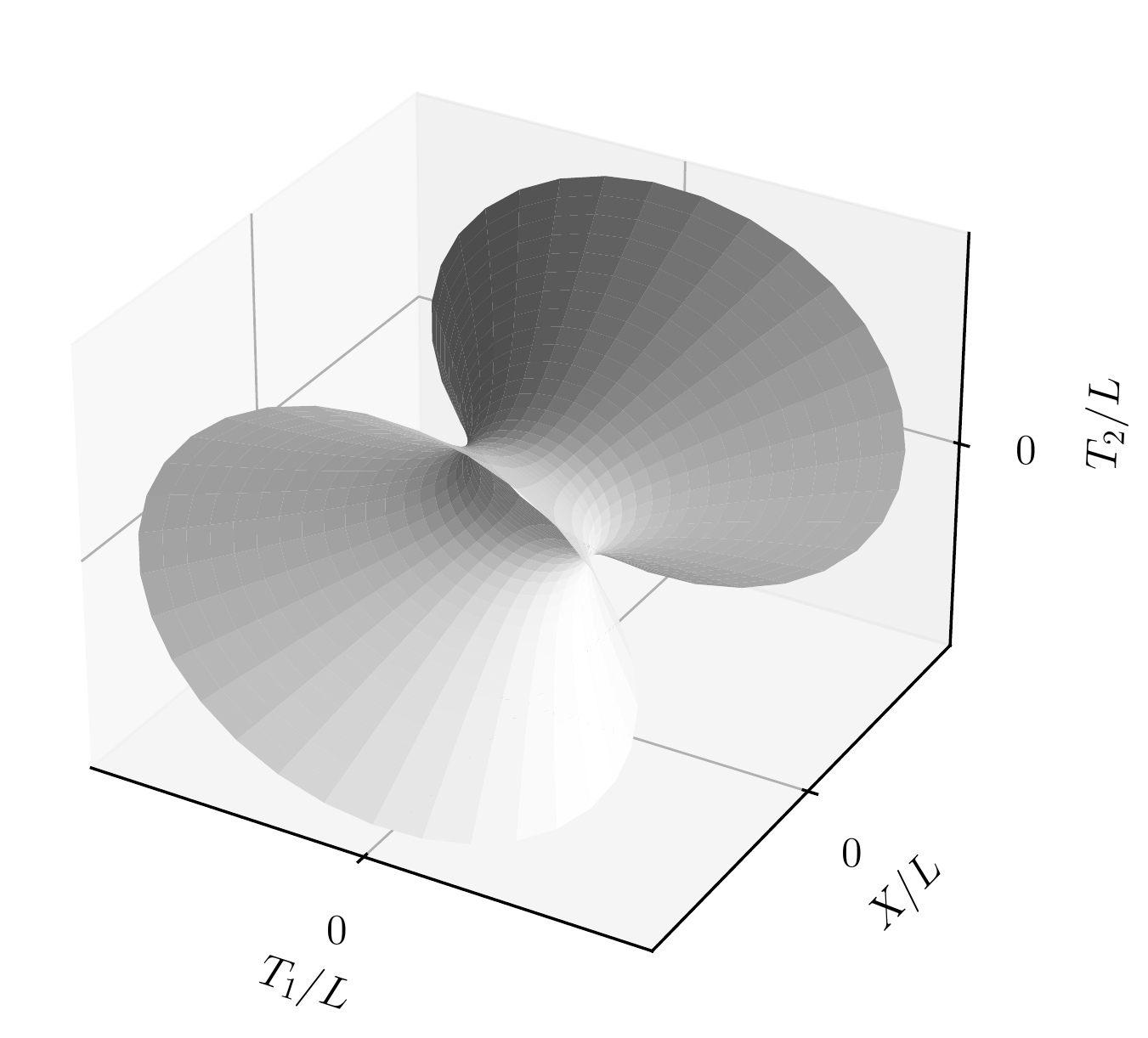} 
	\caption{Embedding of the the $\AdS_2$ in Minkowski space of signature (2,1). The explicit relation between the coordinate is summarized in Eq.~\ref{eqn:Poincare_embedding}. In these coordinates, the two AdS$_2$ boundaries (related by the reflection symmetry) are at $X\to\pm\infty$.  In \Poincare coordinates, these boundaries correspond to $z = 0^\pm$ shown in Fig.~\ref{fig:AdS_causal_diamond}.}\label{fig:hyperboloid}
\end{figure}

The $\AdS_2$ space can be described as a hypersurface $T_1^2+T_2^2-X^2=L^2$ in the Minkowski spacetime with signature (2,1):
\begin{equation}\label{eqn:Minkowski_embedding}
	ds^2 = -dT_1^2 - dT_2^2 + dX^2 \, ,
\end{equation}
As shown in Fig.~\ref{fig:hyperboloid}, this hypersurface is a hyperboloid with one connected component and a reflection symmetry around $X=0$. The two AdS boundaries are located at $X\to\pm\infty$. We see that the boundaries are disjoint and each associated to a coordinate patch. 
The most general solution to Eqs.~\eqref{eqn:classical_EOM} has the dilaton profile $\Phi=AT_1+BT_2+CX$, with some constants $A$, $B$ and $C$. Following Ref.~\cite{harlow2019factorization}, by invoking the SO(2,1) symmetry of the ambient Minkowski spacetime, we can rotate our coordinates such that $B=C=0$. Hence we can write 
\begin{equation}
	\Phi = \frac{\Phi_hT_1}{L},
\end{equation}
where $\Phi_h$ will later be identified as the dilaton value at the horizon. 
\begin{figure}
	\includegraphics[width=1.\textwidth]{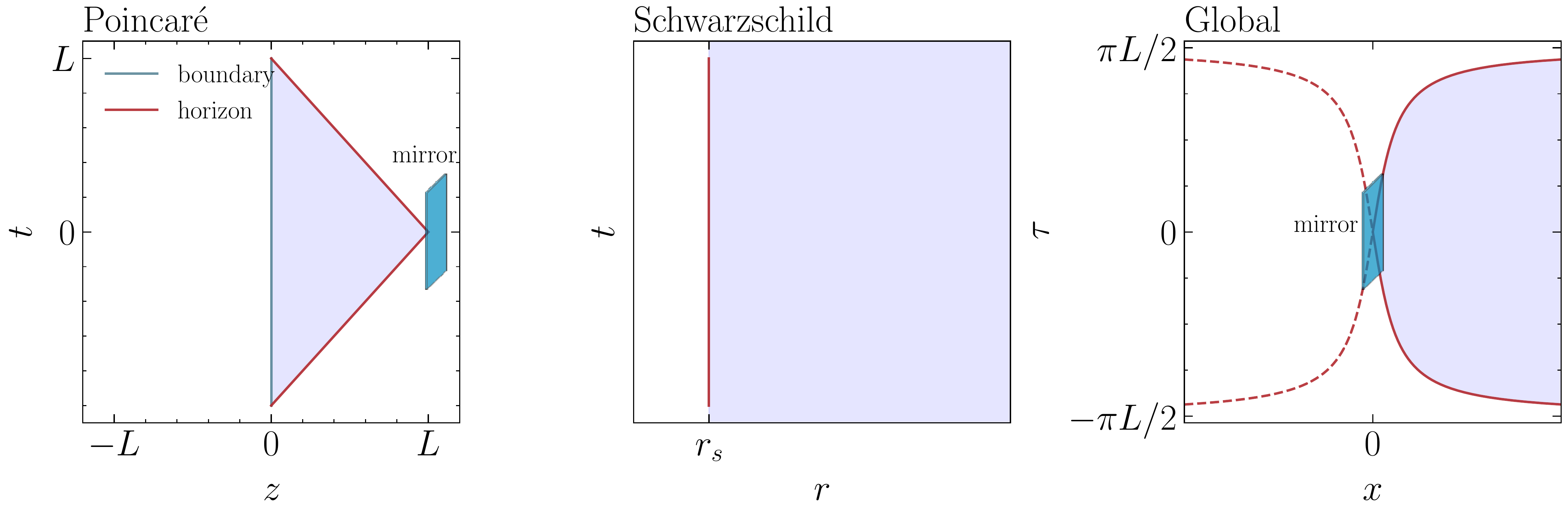} 
	\caption{Causal diamonds in different AdS coordinates.  In all three panels, the shaded region corresponds to a causal diamond in one half of the entire AdS space, which in embedding coordinates is $X \to + \infty$ shown in Fig.~\ref{fig:hyperboloid}.  It is also the shaded region that corresponds to the interior of the causal diamond in the original Minkowski spacetime, which will be the focus of our attention. The causal diamond horizon inherited from the 4-d Minkowski spacetime is indicated as a solid red curve, while the same horizon but on the other side of AdS is indicated as a dashed red curve.}\label{fig:AdS_causal_diamond}
\end{figure}

As shown in Fig.~\ref{fig:AdS_causal_diamond}, we will use multiple coordinates to describe the causal diamond in AdS geometry, as we discuss in detail next.  We cast the embedding coordinates first in the standard \Poincare coordinates.  Then we discuss Schwarzschild coordinates, which have the advantage of making the position of the Rindler horizon explicit, which in turn will be directly related to the value of the dilaton; these coordinates cover only the interior of the causal diamond shown in Fig.~\ref{fig:AdS_causal_diamond}.  Finally, we transform to global coordinates, which cover the entire spacetime and will be the coordinate system of choice for computing the Hartle-Hawking wavefunction and partition function. We will also explain how each of these coordinate systems relates to the coordinates in the original Minkowski metric, where actual measurements are supposed to take place.

\subsection{\Poincare coordinates}
The most commonly used coordinate system in the AdS spacetime is the aforementioned \Poincare coordinate system, related to the embedding coordinates by
\begin{align}\label{eqn:Poincare_embedding}
	T_1 &= \frac{L^2-t^2+z^2}{2z} \nonumber \\
	T_2 &= L\frac{t}{z} \nonumber \\
	X &= \frac{L^2+t^2-z^2}{2z} \, ,
\end{align}
with the metric and dilaton
\begin{align}\label{eqn:Poincare_coodinates}
	d\tilde{s}^2 &= L^2\frac{-dt^2+dz^2}{z^2} \nonumber \\
	\Phi &= \Phi_h\frac{L^2-t^2+z^2}{2Lz} \, .
\end{align}
The AdS boundaries are located at $z=0^{\pm}$, shown in the left-hand panel of Fig.~\ref{fig:AdS_causal_diamond}; they corresponds to $X \to \pm \infty$ in the embedding coordinates, shown in Fig.~\ref{fig:hyperboloid}. 

{}From Eq.~\eqref{eqn:AdS2_S2}, the $(t,z)$ coordinates in AdS directly translate to $(t,r)$ in Minkowski spacetime, such that the horizon in the original Minkowski spacetime is now at $|t|+z=L$. The shaded regions in Fig.~\ref{fig:AdS_causal_diamond} thus correspond to the interior of the causal diamond also in the original Minkowski spacetime.  

\subsection{Schwarzschild Coordinates}

Since we are interested in the behavior near the horizon, a convenient coordinate system is the so-called ``topological black hole,'' or Schwarzschild system of coordinate $(\mathfrak{t},\mr)$, given by
\begin{align}\label{eqn:Schwarzschild_embedding}
	T_1 &= L\frac{\mr}{\mr_s} \nonumber \\
	T_2 &= L\sinh\left(\frac{\mr_s\mathfrak{t}}{L^2}\right)\sqrt{\frac{\mr^2}{\mr_s^2}-1} \nonumber \\
	X &= \pm L\cosh\left(\frac{\mr_s\mathfrak{t}}{L^2}\right)\sqrt{\frac{\mr^2}{\mr_s^2}-1} \, ,
\end{align}
where
\begin{align}\label{eqn:solutions_Sch}
	d\tilde{s}^2 &= -\frac{\mr^2-\mr_s^2}{L^2}d\mathfrak{t}^2 + \frac{L^2}{\mr^2-\mr_s^2}d\mr^2 \nonumber \\
	\Phi &= \Phi_b \frac{\mr}{L} \, ,
\end{align}
and $\mr_s$ is some constant and the coordinate is only defined for $\mr\geq \mr_s$. The $\pm$ sign in Eq.~\eqref{eqn:Schwarzschild_embedding} corresponds to the right and left patches of the AdS spacetime. This coordinate system was used in Refs.~\cite{Casini_2011,vz2} to study the behavior of light sheet horizons utilizing black hole thermodynamics in the bulk. The relation between \Poincare and Schwarzschild coordinates is
\begin{align}\label{eqn:Poincare_Schwarzschild}
	\left(\frac{(L+t)^2-z^2}{2z}\right)\left(\frac{(L-t)^2-z^2}{2z}\right) &= L^2\left(\frac{\mr^2}{\mr_s^2}-1\right) \nonumber \\
	\frac{2Lt}{L^2+t^2-z^2} &= \pm\tanh\frac{\mr_s\mathfrak{t}}{L^2} \, .
\end{align}
It is clear from Eq.~\eqref{eqn:Poincare_Schwarzschild} that $\mr=\mr_s$ is the position of the Rindler (bifurcate) horizon, where $X = T_2 = 0$ corresponds to $t = 0, z = L$ in \Poincare coordinates. The AdS boundary is located at $\mr\to\infty$, hence the region $\mr\geq\mr_s$ describes the entirety of the causal diamond interior.  

Note that Eq.~\eqref{eqn:solutions_Sch} explicitly states that the dilaton controls the position of the Rindler horizon, and evaluating it at the horizon reveals that  
\begin{equation}\label{eqn:r_s}
	\mr_s = L\frac{\Phi_h}{\Phi_b}.
\end{equation}
We thus expect dilaton quantum fluctuations to be responsible for the quantum uncertainty in the photon travel time in the original Minkowski theory.

\subsection{Global Coordinates}
Finally, we define a global coordinate system $(\tau,x)$ by
\begin{align}\label{eqn:global_embedding}
	T_1 &= L\sqrt{1+\frac{x^2}{L^2}}\cos\frac{\tau}{L} \nonumber \\
	T_2 &= L\sqrt{1+\frac{x^2}{L^2}}\sin\frac{\tau}{L} \nonumber \\
	X &= x \, ,
\end{align}
such that
\begin{align}\label{eqn:solutions_global}
	d\tilde{s}^2 &= -\left(1+\frac{x^2}{L^2}\right) d\tau^2 + \frac{dx^2}{1+x^2/L^2} \nonumber \\
	\Phi &= \Phi_h\sqrt{1+\frac{x^2}{L^2}}\cos\frac{\tau}{L} \, ,
\end{align}
Following Ref.~\cite{harlow2019factorization}, this is the basis of choice for computing the Hartle-Hawking wavefunctions and the partition function. An important observation is that the global coordinates cover the entire AdS spacetime, while the \Poincare and (right) Schwarzschild coordinates only cover the region $x\geq 0$, i.e. the right exterior region. This can be easily verified by noting the relation with the \Poincare coordinates
\begin{align}\label{eqn:Poincare_global}
	\tan\frac{\tau}{L} &= \frac{2Lt}{L^2-t^2+z^2} \nonumber \\
	x &= \frac{L^2+t^2-z^2}{2z} \, .
\end{align}
Moreover, one could check that the horizon is located at $x=\pm L\tan{(\tau/L)}$, while the AdS boundary is at $x\to\pm\infty$. Hence, the causal diamond is a subset of the right coordinate patch, while the global coordinates effectively provide the maximal extension of the patch. An analogous coordinate system can be set up to describe the left exterior region, thus effectively factorizing the system. 

With the groundwork laid on the relation between the dilaton and coordinate systems, we can now compute the quantum fluctuations.

%-------------------------------------------------------------------------------------------------
%
% 	         Spacetime Fluctuation in JT Gravity
% %-------------------------------------------------------------------------------------------------

\section{Spacetime Fluctuations in JT Gravity}
\label{sec:spacetime_fluc_JT}

Our analysis mostly follows Ref.~\cite{harlow2019factorization}, which was originally motivated by the factorization problem~\cite{harlow_2016, guica_2017}. Instead of applications to the factorization problem, we use this framework for constructing the action and its solutions beyond the classical saddle point approximation.

One important feature of the JT gravity is that it can be reduced to a 1-d quantum mechanics on the boundary.  The Hamiltonian of the QM problem is obtained by evaluating the stress-energy tensor on each boundary, left and right, using the action in Eq.~\eqref{eqn:JT_def}:
\begin{equation}\label{eqn:H_LR}
	H_L = H_R = \frac{\Phi_h^2}{L\Phi_b} \, .
\end{equation}
The Hamiltonian on the left (resp. right) boundary is conjugate to the time variable $t_L$ (resp. $t_R$), denoting the Schwarschild time on the respective AdS boundary.  Alternatively, on can define conjugate momentum $P$ and length (which we denote $L_g$ to distinguish it from the AdS radius). In these variables, the symplectic form $\Omega$ looks like~\cite{harlow2019factorization}
\begin{equation}\label{eqn:symplectic_form}
	\Omega = d\delta\wedge dH = dL_g\wedge dP \, ,
\end{equation}
where $H=H_L+H_R$ is the total Hamiltonian. The two canonical conjugate pairs are $(\delta,H)$ and $(L_g, P)$.

\subsection{Canonical Variables}
Since Eq.~\eqref{eqn:H_LR} implies that $H_L-H_R=0$, the only time variable is generated by $H_L+H_R$, and is defined to be
\begin{equation}\label{eqn:delta_tLR}
	\delta = \frac{\mathfrak{t}_L+\mathfrak{t}_R}{2} \, .
\end{equation}
It is noted in Ref.~\cite{harlow2019factorization} that $\delta$ can be interpreted as a time-shift operator of the Hilbert space spanned by normalized states $\ket{E}$, with $\delta = i\partial/\partial E$. Physically, $\delta$ is the {\em time difference} between the two boundaries, which is a quantity that can be measured by an interferometer system. According to Ref.~\cite{harlow2019factorization}, one way to define $\delta$ is to examine a geodesic connecting the two boundaries which is orthogonal to surfaces of constant $\Phi$. The fluctuation of the arrival time relative to the starting time is characterized by $2\delta$~\cite{harlow2019factorization}. A suitable candidate for such geodesic is simply the horizon of the two-sided AdS, defined by firing a photon from a point at the left boundary, $(-\pi L/2,-\infty)$, and eventually arriving at the right boundary, $(\pi L/2,\infty)$. The horizon is indicated as a red line in the right panel of Fig.~\ref{fig:AdS_causal_diamond} (ignoring the mirror in the figure for now), combining the dashed and the solid lines in the left and right AdS patch respectively. The equation for this trajectory can be solved by setting $d\tilde{s}=0$ in Eq.~\eqref{eqn:Poincare_global}, which turns out to be $\tau = L\tan^{-1}(x/L)$\footnote{Usually it is not possible to shoot a photon from one boundary to the other in a two-sided black hole system, since the two boundaries are causally disconnected. However, in our set-up, the photon trajectory defines the horizon, which is a (and the only) null geodesic that connects the two boundaries, as apparent when the Penrose diagram of the spacetime is inspected. AdS geometries with horizons defined by photon paths are also noted and used in Refs~\cite{Casini_2011, vz2}.}. The dilaton field along this trajectory can be found by putting this equation into  Eq.~\eqref{eqn:Poincare_global}, which turns out to be a constant, $\Phi=\Phi_h$, as expected. Since null geodesics are orthogonal to themselves, the horizon (which connects the two boundaries) satisfies the condition quoted from Ref.~\cite{harlow2019factorization}, {\em i.e.} a geodesic that is orthogonal to curves of constant $\Phi$. Hence we can interpret $\delta$ as the relative time between the two boundaries measured by this particular photon path.  Note that $\delta = 0$ corresponds to the classical (unperturbed) light trajectory, since the clock on the left and right boundaries runs oppositely, and the time for a photon to traverse from $x = 0$ to either boundary is the same in the unperturbed spacetime, {\em i.e.} the clocks on each boundary tick at the same pace (but with opposite arrows of time) on either boundary in the absence of quantum fluctuations.  {\em Thus $\delta \neq 0$ indicates a quantum fluctuation in the light trajectory, or equivalently, a quantum fluctuation in rate at which the boundary clocks tick.}

\begin{figure}
	\includegraphics[scale=0.8]{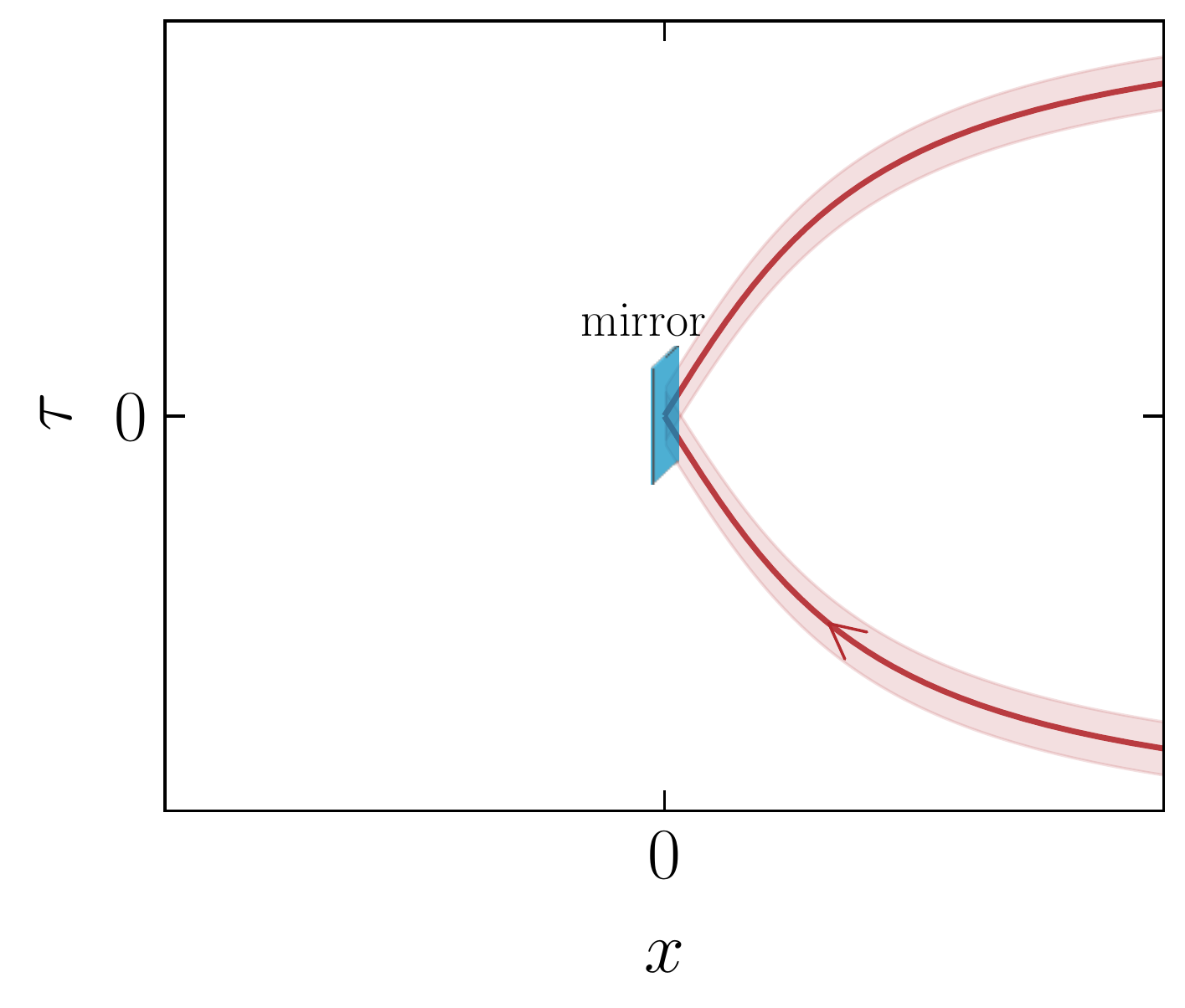} 
	\caption{The quantum uncertainty in the light trajectory, here depicted by fuzzing of the horizon, is what we seek to compute via the quantum uncertainty in the geodesic distances parameterized by $L_g$ and $\delta$ defined in the text.  In particular $2 \delta$ is the time shift, with respect to a classical unperturbed trajectory, for a photon that is fired from the right boundary and reflected back to its starting position.}\label{fig:geodesic_fuzzing}
\end{figure}

Further, $\delta$ is related to the quantum fluctuation in the time of arrival of a photon in the Minkowski interferometer. First, we note that while the two-sided AdS system is a natural solution to the JT theory, the original Minkowski causal diamond only covers a \Poincare patch as indicated by the shaded regions in Fig.~\ref{fig:AdS_causal_diamond}. However, by putting a mirror at $x=0$ ({\em i.e.} the interface between the two AdS sides), one can construct a geodesic that was fired from the right boundary at $(-\pi L/2,\infty)$, reflected by the mirror at $(0,0)$, and arrives back at the right boundary at $(\pi L/2,\infty)$, as indicated by the solid red line at the right panel of Fig.~\ref{fig:AdS_causal_diamond}. This is simply the horizon of the Minkowski causal diamond. Then, using reflection symmetry around $x=0$ as discussed in the previous paragraph, the distance traveled by this photon must be identical, in the absence of quantum fluctuations, to the distance traveled by a photon fired from the left boundary and eventually arrives at the right boundary, {\em i.e.} the two-sided AdS horizon. Then, $2\delta$ is precisely the time shift, with respect to a classical unperturbed trajectory, for a photon that is fired from the right boundary and reflected back to its starting position. This is illustrated in Fig.~\ref{fig:geodesic_fuzzing}.

In this sense,  the two-sided AdS serves as a mathematical trick (philosophically akin to the method of images in electrostatics) for us to compute the physical photon travel time, by allowing us to compute the relative photon travel time in one copy of AdS  with respect to the other.  Since the photon travel time must be the same on both sides in the absence of quantum fluctuations, $\delta$ thus quantifies quantum fluctuations in the time of arrival of the photon in one copy of AdS relative to the reference copy.  

We further work with the assumption that the mirror is a probe and hence does not substantially affect the spacetime geometry. This is analogous to calculations of the interferometer response in gravitational wave experiments, which also neglect the back-reaction of the geometry to the mirrors. This treatment can be justified by considering the Schwarzschild radius of the mirror. For a mirror with mass $\sim10$~kg, its Schwarzschild radius is $R_{\mathrm{mirror}}=2G_NM_{\mathrm{mirror}}\sim10^{-26}$~m, which is much shorter than both the interferometer arm length and the photon wavelength, and hence its back-reaction to the geometry can be ignored. On the other hand, since $R_{\mathrm{mirror}}$ is much longer than $l_p$, we can also ignore the quantum effect of the mirror. We believe that the effect of the mirror could be included more explicitly by incorporating the additional degrees of freedom associated with the reflecting boundary conditions (a la ``end-of-the-world brane''), but we leave this implementation to future work.  

Now we turn to the other pair of canonically conjugate variables $(L_g, P)$.  The renormalized geodesic distance $L_g$ between the two boundaries can be evaluated using the global coordinates in Eq.~\eqref{eqn:solutions_global}, where the boundaries are now regulated by bringing them from $x \rightarrow \pm \infty$ to some cut-off at $x=\pm x_c$. The expression for $x_c$ can be found by equating the second lines of Eq.~\eqref{eqn:solutions_Sch} and Eq.~\eqref{eqn:solutions_global} at the boundary
\begin{align}\label{eqn:xc}
	\Phi|_{\partial\tilde{M}_2} = \Phi_b\frac{\mr_c}{L} &= \Phi_h\sqrt{1+\frac{x_c^2}{L^2}}\cos\frac{\tau}{L} \nonumber \\
	\implies x_c &\approx \frac{L\mr_c}{\mr_s}\frac{1}{\cos(\tau/L)} \, ,
\end{align}
where we used Eq.~\eqref{eqn:r_s} and also assumed $x_c\gg L$. This allows us to define a ``renormalized geodesic distance'' \cite{harlow2019factorization}
\begin{align}\label{eqn:renormalized_geodeisc_distance}
	L_g &= \int_{-x_c}^{x_c}\sqrt{-g_{\tau\tau}\frac{d\tau}{dx}\frac{d\tau}{dx}}dx - 2L\log(2\Phi|_{\partial\tilde{M}_2}) \nonumber \\
	&= \int_{-x_c}^{x_c}\frac{1}{\sqrt{1+x^2/L^2}}dx - 2L\log\left(\frac{2\Phi_b\mr_c}{L}\right) \nonumber \\
	&= 2L\sinh^{-1}\left(\frac{x_c}{L}\right)-2L\log\left(\frac{2\Phi_b\mr_c}{L}\right) \nonumber \\
	&\approx 2 L\log \left(\frac{x_c}{\Phi_b\mr_c}\right) \nonumber \\
	&= 2 L\log \left(\frac{L\cosh(\mr_s\delta/L^2)}{\Phi_b\mr_s}\right)\, ,
\end{align}
where we used $\cos(\tau/L)=1/\cosh(\mr_s\delta/L^2)$ in the third line~\cite{harlow2019factorization}, which can be found by equating Eq.~\eqref{eqn:Schwarzschild_embedding} and Eq.~\eqref{eqn:global_embedding} and taking the $\mr_c\gg \mr_s$ limit.  We are interested in computing the Euclidean path integral in terms of $L_g$, and because we are interested in perturbations about the classical spacetime (where $\delta = 0$), we will expand $L_g$ to its first correction in $\delta$:
\begin{equation}\label{eqn:L_delta}
	L_g \approx 2L\left(\log \frac{L}{\Phi_b r_s} + \frac{\mr_s^2 \delta^2}{2L^4}\right).
\end{equation}
That $L_g$ depends (at first order) quadratically on $\delta$ will have important consequences for fluctuations in the photon travel time.

\subsection{Euclidean Path Integral}
We now turn to the solution to the QM path integral from a saddle point expansion of the Euclidean Path Integral, which gives the thermodynamic fluctuations of the system. 
The saddle-point geometry in Euclidean signature is given by performing a Wick rotation on Eq.~\eqref{eqn:solutions_Sch}
\begin{align}\label{eqn:solutions_Sch_euclidean}
	d\tilde{s}_E^2 &= \frac{\mr^2-\mr_s^2}{L^2}d\mathfrak{t}_E^2 + \frac{L^2}{\mr^2-\mr_s^2}d\mr^2 \, .
\end{align}
Consider the AdS geometry with one asymptotic boundary. To avoid a conical singularity at $\mr=\mr_s$, we require
\begin{align}\label{eqn:rs_2pi_beta}
	\mr_s = \frac{2\pi L^2}{\beta} \, .
\end{align}
We compute the JT action in Eq.~\eqref{eqn:JT_def} on an AdS manifold with a disk topology and $\beta$ as the periodicity of $t_E$,
\begin{equation}\label{eqn:JT_euclidean}
	-I_E = 2\frac{\Phi_b\mr_c}{L}\int_0^\beta d\mathfrak{t}_E \sqrt{\tilde{\gamma}_1}\tilde{K}_1 \, ,
\end{equation}
where we have used the bulk equation of motion and the dilaton value at the boundary. The Euclidan version of the boundary condition outlined below Eq.~\eqref{eqn:classical_EOM} is $\sqrt{\tilde{\gamma}}=\sqrt{\tilde{\gamma}_{t_Et_E}}=\mr_c/L$. The unit vector normal to the boundary $\mr=\mr_c$ is $\tilde{n}^{\mu}=(0,\sqrt{\mr^2-\mr_s^2}/L)$. Hence, the extrinsic curvature in Eq.~\eqref{eqn:solutions_Sch_euclidean} is given by
\begin{align}\label{eqn:extrinsic}
	\tilde{K}_1 &= \dt_{\mu}\tilde{n}^{\mu}|_{\mr=\mr_c} \nonumber \\
	&= \partial_{\mu}\tilde{n}^{\mu}|_{\mr=\mr_c} \nonumber \\
	&= \frac{1}{L}\frac{\mr_c}{\sqrt{\mr_c^2-\mr_s^2}} \nonumber \\
	&= \frac{1}{L} \left(1+\frac{1}{2}\frac{\mr_s^2}{\mr_c^2}\right) \, ,
\end{align}
where we used $\sqrt{\gt}=1$ in the second line. Finally, putting Eq.~\eqref{eqn:extrinsic} and Eq.~\eqref{eqn:rs_2pi_beta} into Eq.~\eqref{eqn:JT_euclidean}, the action becomes\footnote{To obtain a finite result in Eq.~\eqref{eqn:JT_euclidean}, we need to add a holographic renormalization counterterm $-(2/L)\int_{\partial\tilde{M}_2} dx^0 \sqrt{-\tilde{\gamma}_1} \Phi$, similar to the one in Ref.~\cite{Maldacena:2016upp} where the Schwarzian action is derived from the JT action, but with a different boundary condition.}
\begin{equation}\label{eqn:JT_euclidean_evaluated}
	-I_E = 4\pi^2 L\frac{\Phi_b}{\beta} \, .
\end{equation}
The thermal partition function {\em evaluated at the saddle-point} is given by 
\begin{align}\label{eqn:thermal_partition}
	Z[\beta] &= e^{-I_E} \nonumber \\
	&= e^{4\pi^2L\Phi_b/\beta} \, .
\end{align}
This allows to compute the energy and the entropy 
\begin{align}\label{eqn:energy_entropy}
	\langle E\rangle = -\partial_{\beta}\log Z[\beta] &= \frac{1}{L}\frac{\Phi_h^2}{\Phi_b} \nonumber \\
	S = \log Z[\beta] + \beta \langle E\rangle &= 4\pi\Phi_h \, .
\end{align}
Here we see the direct connection between the entropy and the value of the dilaton at the horizon.

We can get the leading correction to the saddle-point via
\begin{equation}
Z\left[\beta\right] \approx \int_0^\infty dE_L  e^{S(E_L)-\beta E_L} \approx \int_0^\infty dE_L  e^{4\pi \sqrt{L\Phi_b E_L}-\beta E_L}.
\end{equation}
This is the famous ``square-root E'' behavior of the density of states that appears in many systems.  It was shown in Ref.~\cite{banks} that this density of states gives rise to the relation $\beta^2 \partial_\beta^2 \log Z\left[\beta\right]  = -\beta \partial_\beta \log Z\left[\beta\right] + \log Z\left[\beta\right]$, which corresponds to $\langle \Delta K^2 \rangle = \langle K \rangle$ \cite{de_Boer_2019,vz2} in the language of AdS/CFT. This also directly follows from the relation $\log Z\sim\beta^{-1}$ at the saddle-point as indicated in Eq.~\eqref{eqn:thermal_partition}.

We will later identify the entropy of the system to be the black hole entropy associated with the causal diamond horizon, {\em i.e.}
\begin{align}\label{eqn:black_hole_entropy}
	S &= \frac{A}{4G_N} \nonumber \\
	&= \frac{8\pi^2L^2}{l_p^2} \, .
\end{align} 

In order to understand the fluctuation in $\delta$, we now turn our attention to the calculation in the $(L_g, P)$ basis with two asymptotic boundaries in global coordinates. Following Ref.~\cite{harlow2019factorization}, this can be achieved by studying the Hartle-Hawking wavefunction, which can be
interpreted as a wormhole connecting the two boundaries. Operationally, this amounts to computing the action in Eq.~\eqref{eqn:JT_def} with the metric in Eq.~\eqref{eqn:solutions_Sch_euclidean}, where the boundaries of the manifold is now the AdS conformal boundary with length $\mr_c\beta/2$ and a bulk boundary $\Sigma$. The action in Eq.~\eqref{eqn:JT_def} also has to be modified to include contributions from the two corners of the geometry. The result is
\begin{equation}\label{eqn:Euclidean_wavefunction}
	-I_E = \frac{8L\Phi_b}{\beta}\left(y^2+\frac{2y}{\tan y}\right) \, ,
\end{equation} 
where $\beta$ is the periodicity of the Euclidean time and
\begin{align}\label{eqn:a_x}
	y &= \frac{r_s\beta}{4L^2} = \frac{1}{4}\frac{\beta\Phi_h}{L\Phi_b} \nonumber \\
	a &= \frac{\sin y}{y} = 4L\Phi_be^{L_g/2L}\beta^{-1} \, , 
\end{align} 
and $a\leq 1$. We observe that $I_E$ is minimized at $y=\pi/2$, which corresponds to $\delta=0$ according to Eq.~\eqref{eqn:renormalized_geodeisc_distance}. Expanding near the peak, one finds~\cite{harlow2019factorization}
\begin{align}\label{eqn:action_expansion_tau}
	-I_E &= \mathrm{constant} - \frac{8L\Phi_b}{\beta}\left(y-\frac{\pi}{2}\right)^2 \nonumber \\
	&= \mathrm{constant} - \frac{\pi^2}{2}\frac{\Phi_b}{\beta L}(L_g-L_{g,\mathrm{peak}})^2 \nonumber \\
	&= \mathrm{constant} - \frac{S}{16L^2}(L_g-L_{g,\mathrm{peak}})^2 \, ,
\end{align} 
where in addition to Eq.~\eqref{eqn:energy_entropy}, we used Eq.~\eqref{eqn:rs_2pi_beta} in the last line, which is expected to hold at the peak of the wavefunction as required by smoothness at $\mr=\mr_s$ in Eq.~\eqref{eqn:solutions_Sch}. This suggests that the uncertainty of $L_g$ is
\begin{equation}\label{eqn:Delta_L}
	\Delta L_g = \frac{2\sqrt{2}L}{\sqrt{S}} \, ,
\end{equation} 
Using Eq.~\eqref{eqn:L_delta}, this translates to the variance in $\delta$
\begin{align}\label{eqn:variance_delta}
	\Delta \delta^2 = \frac{2\sqrt{2}L^4}{\mr_s^2\sqrt{S}} \, . 
\end{align} 
We note that the precise numerical factor here depends on the details of the path integral measure, which we mostly ignored so far in our leading-order analysis. Moreover, at this level of approximation, we can use semiclassical relations between different variables, in particular, between $L_g$ and $\delta$. A more careful treatment would require a Jacobian factor in the path integral, which also can be considered as a part of the integration measure. We expect that all such factors do not considerably change the results of the leading-order saddle point analysis.

%-------------------------------------------------------------------------------------------------
%
% 	         Photon Travel time
% %-------------------------------------------------------------------------------------------------

\section{Photon Travel Time}
\label{sec:photon_travel_time}

\begin{figure}
	\includegraphics[scale=0.8]{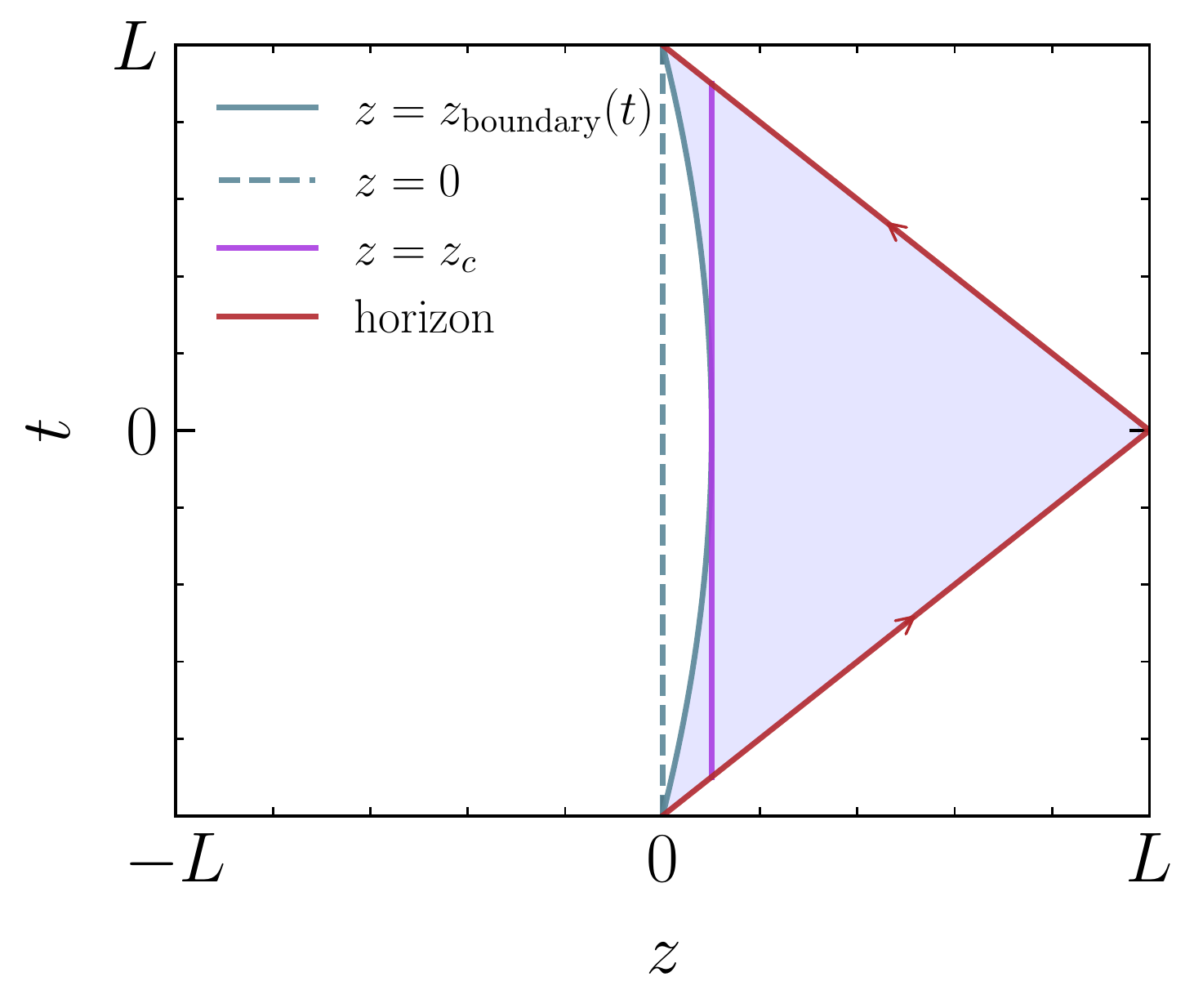} 
	\caption{Spacetime diagram showing the regularized boundary $z=z_{\mathrm{boundary}}(t)$, the flat brane $z=z_c$ and the photon trajectories. We have chosen $\mr_s/\mr_c=0.2$ for illustration purpose.}\label{fig:z_boundary}
\end{figure}

The uncertainty relation in Eq.~\eqref{eqn:action_expansion_tau} allows us to compute the uncertainty in photon travel time in the interferometer system. Recall that $\delta$ measures the time shift between the two AdS boundaries shown in the right-hand panel of Fig.~\ref{fig:AdS_causal_diamond}.  When regulated, the boundaries are brought in from $\mr \rightarrow \infty$ to a finite value $\mr=\mr_c$ in their respective Schwarzschild patch. To allow for a non-zero value of $\delta$, we must allow the two boundaries to fluctuate independently while keeping $\Phi_b$ fixed. Since the experiment is only probing the right exterior region ({\em i.e.} $z>0$), we would like to trace out the degrees of freedom in the left patch. This can be achieved by taking the limit where $\mr_c$ in the left is much greater than its right-hand side counterpart. Operationally, we take $\mr_c\to\infty$ at the left while keeping $\mr_c$ at the right finite (but still large). In \Poincare coordinates, this perturbs the boundary from $z=0^+$ to some small curve $z=z_{\mathrm{boundary}}(t)$. Putting $\mr=\mr_s$ in the first line of Eq.~\eqref{eqn:Poincare_Schwarzschild}, one finds
\begin{align}\label{eqn:z_boundary}
	z_{\mathrm{boudnary}}(t) &= \frac{L-\sqrt{L^2-(L^2-t^2)(\mr_s^2/\mr_c^2)}}{\mr_s/\mr_c} \nonumber \\
	&= \frac{L^2-t^2}{2L}\frac{\mr_s}{\mr_c} + \mathcal{O}\left(\frac{\mr_s^3}{\mr_c^3}\right) \, .
\end{align}
As expected, if $\mr_c\to\infty$, the boundary would be located at $z=0$. The regularized boundary $z=z_{\mathrm{boundary}}(t)$ turns out to be a parabola, which we plot in Fig.~\ref{fig:z_boundary}.
Noting that the left boundary is still at $z=0$, the boundary times are given by the second line of Eq.~\eqref{eqn:Poincare_Schwarzschild}
\begin{align}\label{eqn:t_LR}
	\mathfrak{t}_R &= \frac{L^2}{\mr_s}\tanh^{-1}\frac{2Lt}{L^2+t^2-z^2_{\mathrm{boundary}}(t)} \nonumber \\
	\mathfrak{t}_L &= -\frac{L^2}{\mr_s}\tanh^{-1}\frac{2Lt}{L^2+t^2} \, .
\end{align}
We emphasize that $t$ is the \Poincare time, related to the Minkowski time by a conformal rescaling (as already noted below Eq.~\eqref{eqn:AdS2_S2}), while $\mathfrak{t}$ is Schwarzschild time. Hence, we use Eq.~\eqref{eqn:z_boundary} to evaluate $\delta=(\mathfrak{t}_L+\mathfrak{t}_R)/2$ and obtain
\begin{equation}\label{eqn:delta}
	\delta = \frac{L\mr_s}{4\mr_c^2}t + \mathcal{O}\left(\frac{\mr_s^3}{\mr_c^3}\right) \, .
\end{equation}
 We see that $\delta$ in linear in the \Poincare time. 
 
The interferometer is now placed on a flat brane $z=z_c$ for some constant $z_c\ll L$ such that the brane is barely touching $z_{\mathrm{boundary}}(t)$. This ensures that the brane is as close to the boundary as possible without leaving the domain of the system, which sets the value of $z_c$ to be
\begin{equation}\label{eqn:zc}
	z_c = \frac{L}{2}\frac{\mr_s}{\mr_c} \, .
\end{equation}
The location of the brane is also indicated in Fig.~\ref{fig:z_boundary}. The photon round-trip time $T_{\RT}$ is $2(L-z_c)$ multiplied by a conformal factor in front of the metric $L/z_c$, which is approximately
\begin{align}\label{eqn:roundtrip}
	T_{\RT} &= \frac{2L^2}{z_c} \nonumber \\
	&= 4L\frac{\mr_c}{\mr_s} \, .
\end{align}
Note that the photon travel time diverges if the boundary was not regularized, as noted in Ref.~\cite{vz2}. The fluctuation in photon roundtrip time scales linearly with fluctuations in $\delta$.
The ratio $\Delta T_{\RT}/T_{\RT}$ should be independent of the metric prefactor. Using Eq.~\eqref{eqn:delta}, Eq.~\eqref{eqn:variance_delta} and Eq.~\eqref{eqn:roundtrip}, we find
\begin{align}\label{eqn:RT_ratio}
	\frac{\Delta T_{\RT}^2}{T_{\RT}^2} &=  \frac{(4\mr_c^2/L\mr_s)^2\Delta \delta^2}{4L^2} \nonumber \\
	&= 8\sqrt{2}\left(\frac{\mr_c}{\mr_s}\right)^4\frac{1}{\sqrt{S}} \nonumber \\
	&= \frac{1}{\sqrt{2}}\left(\frac{T_{\RT}}{2L}\right)^4\frac{1}{\sqrt{S}} \, .
\end{align}
Since the experiment is carried out in Minkowski spacetime, the photon measured travel time does not have any conformal factors in it, which allow us to identify $T_{\RT}=2L$. Combined with the entropy relation in Eq.~\eqref{eqn:black_hole_entropy}, we find
\begin{align}\label{eqn:RT_final}
	\frac{\Delta T_{\RT}^2}{T_{\RT}^2} &= \frac{1}{\sqrt{2S}} \nonumber \\
	&= \frac{l_p}{4\pi L} \, .
\end{align}
This scaling relation agrees with the previous work of one of the present authors in Refs.~\cite{vz1, vz2, zurek2021vacuum, banks}, which demonstrated that the two-point correlation function of arm length fluctuations in an interferometer system are proportional to $l_p/L$.  While a small fluctuation, it is measurable with a laboratory scale interferometer.

%-------------------------------------------------------------------------------------------------
%
% 	         Conclusion
% %-------------------------------------------------------------------------------------------------

\section{Conclusion}
\label{sec:conclusion}

The dimensional reduction of Einstein gravity in a causal diamond of the four-dimensional flat Minkowski spacetime can be described by two-dimensional JT gravity with the dilaton field playing the role of the diamond area. By analyzing the Hartle-Hawking wavefunction in JT gravity, we find that the uncertainty in an interferometer arm length scales with the Newton's constant as $\Delta L\sim\sqrt{l_pL}$. This agrees with the previous works~\cite{vz1, vz2, zurek2021vacuum, banks}, where the same scaling was obtained by other methods.

Our result in Sec.~\ref{sec:spacetime_fluc_JT} may appear surprising since it naively violates well-known effective field theory lore, which states that the two-point function of an observable should scale with an integer power of the coupling constant. Our result, however, is not in contradiction with this fact for two reasons. First, our analysis is not based on perturbation theory involving a single graviton. Instead, this is a collective effect that comes from all quantum gravity effects within a causal diamond. This is analogous to hydrodynamic description of diffusion, where the UV scale is the average separation of fluid particles. In hydrodynamics, it is well-known that a particle in the system admits a random walk description, with variance growing linearly in time and with the diffusion coefficient that scales as the square root of the UV scale. Following Refs.~\cite{nickel, crossley_2016}, relations between JT gravity and hydrodynamics have been noted {\em e.g.} in Ref.~\cite{harlow2019factorization}. Establishing a more precise connection between quantum gravity in flat spacetime and hydrodynamics is a possible future development of this work.

The second reason our result is consistent with the effective field theory lore is that the quantity with a traditional EFT scaling $\langle L_gL_g\rangle\sim S^{-1}\sim G_N$ does not correspond to the observable $\delta$ relevant for a photon travel time measurement.  Rather $\delta$ scales as $L_g \sim \delta^2$, implying that it is the four-point of $\delta$ with linear scaling in $G_N$.  This behavior is familiar from the study of time-ordered/out-of-time-ordered-correlators, and it is not surprising that such correlators are relevant for systems that display chaotic and hydrodynamic behavior.  We leave study of the connection between the observable of interest and hydrodynamic and chaotic behavior for future work.

\acknowledgments

We thank Tom Banks, Temple He, Cynthia Keeler, Juan Maldacena, Allic Sivaramakrishnan and Erik Verlinde for discussion on these directions. KZ and VL are supported by the Heising-Simons Foundation ``Observational Signatures of Quantum Gravity'' collaboration grant 2021-2817, and by a Simons Investigator award.  The work of SG and KZ~is supported by the U.S.~Department of Energy, Office of Science, Office of High Energy Physics, under Award No.~DE-SC0011632.

\appendix

% %-------------------------------------------------------------------------------------------------
% 	         Bibliography
% %-------------------------------------------------------------------------------------------------

\bibliography{bibliography}

\end{document}